\documentclass[preprint,aps,showpacs,preprintnumbers,amsmath,superscriptaddress,floatfix,amssymb]{revtex4}

\usepackage{graphicx}
\usepackage{dcolumn}
\usepackage{bm}

\begin{document}

\preprint{}

\title{Decay modes of $^{10}$C nuclei unbound state}


\author{F.~Grenier}
\affiliation{ Laboratoire de Physique Nucl\'eaire, Universit\'e Laval,
Qu\'ebec, Canada G1K 7P4.}
\author{A.~Chbihi}
\affiliation{ GANIL, CEA et IN2P3-CNRS, B.P.~5027, F-14076 Caen Cedex, France.}
\author{R.~Roy}
\affiliation{ Laboratoire de Physique Nucl\'eaire, Universit\'e Laval,
Qu\'ebec, Canada G1K 7P4.}
\author{G.~Verde}
\affiliation{Istituto Nazionale di Fisica Nucleare, Sezione di Catania, I-95123, Catania, Italy}
\affiliation{ GANIL, CEA et IN2P3-CNRS, B.P.~5027, F-14076 Caen Cedex, France.}
\author{D. Th\'{e}riault}
\affiliation{ Laboratoire de Physique Nucl\'eaire, Universit\'e Laval,
Qu\'ebec, Canada G1K 7P4.}
\author{J.D.~Frankland}
\affiliation{ GANIL, CEA et IN2P3-CNRS, B.P.~5027, F-14076 Caen Cedex, France.}
\author{J.P.~Wieleczko}
\affiliation{ GANIL, CEA et IN2P3-CNRS, B.P.~5027, F-14076 Caen Cedex, France.}
\author{B.~Borderie}
\affiliation{ Institut de Physique Nucl\'eaire, IN2P3-CNRS, F-91406 
Orsay Cedex,  France.}
\author{R.~Bougault}
\affiliation{ LPC, IN2P3-CNRS, ISMRA et Universit\'e, F-14050 Caen Cedex, France.}
\author{R.~Dayras}
\affiliation{ DAPNIA/SPhN, CEA/Saclay, F-91191 Gif sur Yvette, France.}
\author{E.~Galichet}
\affiliation{ Institut de Physique Nucl\'eaire, IN2P3-CNRS, F-91406 Orsay Cedex,
 France.}
\affiliation{ Conservatoire National des Arts et M\'etiers, F-75141 Paris
Cedex 03.}
\author{D.~Guinet}
\affiliation{ Institut de Physique Nucl\'eaire, IN2P3-CNRS et Universit\'e
F-69622 Villeurbanne, France.}
\affiliation{ Institut de Physique Nucl\'eaire, IN2P3-CNRS, F-91406 Orsay Cedex,
 France.}
\author{P.~Lautesse}
\affiliation{ Institut de Physique Nucl\'eaire, IN2P3-CNRS et Universit\'e
F-69622 Villeurbanne, France.}
\author{N.~Le~Neindre}
\affiliation{ Institut de Physique Nucl\'eaire, IN2P3-CNRS, F-91406 
Orsay Cedex,  France.} 
\author{O.~Lopez}
\affiliation{ LPC, IN2P3-CNRS, ISMRA et Universit\'e, F-14050 Caen Cedex, France.}
\author{J.~Moisan}
\affiliation{ GANIL, CEA et IN2P3-CNRS, B.P.~5027, F-14076 Caen Cedex, France.}
\author{L.~Nalpas}
\affiliation{ DAPNIA/SPhN, CEA/Saclay, F-91191 Gif sur Yvette, France.}
\author{M.~P\^arlog}
\affiliation{ National Institute for Physics and Nuclear Engineering, RO-76900
Bucharest-M\u{a}gurele, Romania.}
\affiliation{ LPC, IN2P3-CNRS, ISMRA et Universit\'e, F-14050 Caen Cedex, France.}
\author{M.~F.~Rivet} 
\affiliation{ Institut de Physique Nucl\'eaire, IN2P3-CNRS, F-91406 
Orsay Cedex,  France.} 
\author{E.~Rosato}
\affiliation{ Dipartimento di Scienze Fisiche e Sezione INFN, Univ. di 
Napoli ``Federico II'', I-80126 Napoli, Italy.}
\author{B.~Tamain}
\affiliation{ LPC, IN2P3-CNRS, ISMRA et Universit\'e, F-14050 Caen Cedex, France.}
\author{E.~Vient}
\affiliation{ LPC, IN2P3-CNRS, ISMRA et Universit\'e, F-14050 Caen Cedex, France.}
\author{M.~Vigilante}
\affiliation{ Dipartimento di Scienze Fisiche e Sezione INFN, Univ. di 
Napoli ``Federico II'', I-80126 Napoli, Italy.}

\collaboration{INDRA collaboration} \noaffiliation

\date{\today}

\begin{abstract}
Unbound states of $^{10}$C nuclei produced as quasi-projectiles
in $^{12}$C+$^{24}$Mg collisions at E/A = 53 and 95 MeV
are studied with the Indra detector array. Multi-particle correlation function analyses provide 
experimental evidence of sequential de-excitation mechanisms through
the production of intermediate $^{9}$B, $^{6}$Be and $^{8}$Be unbound
nuclei. The relative contributions of different decay sequences  to the total decay width of the explored states is estimated semi-quantitatively. The obtained results show that heavy-ion collisions can be used as a tool to access spectroscopic information about exotic nuclei. 
\end{abstract}

\pacs{25.70.Pq, 21.10.-k, 25.70.-z}
\keywords{Multi-particle correlation functions; Sequential decays; Unbound nuclei}
\maketitle

Energetic heavy-ion collisions have been extensively studied to extract
information about the properties of nuclear matter under extreme conditions \cite{wci06}. 
These studies have also shown that a large variety of isotopes is produced
during the dynamical evolution of the reaction. Some of these isotopes, being 
far from the valley of stability, live temporarily and decay by
particle emission. Their internal particle unbound states can then be 
isolated and studied by means of correlation function techniques \cite{han52,han56,koo77,tan04}. As an example, proton-$^{7}$Be correlation functions have been recently measured \cite{tan04} in central Xe+Au collisions at E/A=50 MeV to determine the spin of internal unbound states of the astrophysically important $^{8}$B nucleus. In this respect, a collision between two heavy ions can be viewed not only as a tool to study nuclear dynamics but also as a laboratory to produce several
nuclear species in one single experiment and study their spectroscopic
properties. This aspect of heavy-ion collision experiments certainly represents an important perspective to access information about the properties of very exotic nuclei. 

Among all exotic nuclear species that can be produced in nuclear reactions, $^{10}$C can be considered as an especially interesting system. Antisymmetrized
Molecular Dynamics calculations (AMD) predict a molecular structure
for the ground state of $^{10}$C \cite{kan97}. Exotic features, such as molecular states and clustering \cite{cha88a,cha88b,wuo92,dip99,yam05,fre99,fre01,fre06,voertzen06}, have indeed attracted the interest of a large community and particle correlation analyses have been used to access such features experimentally \cite{wuo92,dip99,yam05,fre99,fre01,fre06,voertzen06}. However, previous studies of $^{10}$C nuclei have hardly been capable
of reconstructing those states lying above the particle emission
threshold of 3.73 MeV \cite{niveau8} and decaying into a single final configuration constituted by two alpha particles and two protons (2$\alpha$+2$p$). All intermediate states that can be formed starting from the decay of $^{10}$C nuclei into charged particle emission (i.e. $^{2}$H, $^{8}$Be, $^{5}$Li, $^{6}$Be, $^{9}$B) are unstable. Reconstructing the decay paths of $^{10}$C states therefore requires detection capabilities with high energy and angular resolution over a large solid angle. 

In this article we access highly lying states in $^{10}$C nuclei produced as excited quasi-projectiles in $^{12}$C+$^{24}$Mg peripheral collisions at E/A=53 and 95 MeV. By means of three- and four-particle correlation
functions we provide experimental evidence of sequential decay modes
for these states through the production of intermediate unbound $^{9}$B,
$^{6}$Be and $^{8}$Be nuclei. Exploring the relative contributions of these sequential decay processes to the total decay widths of $^{10}$C states provides important spectroscopic information that is relevant to access branching ratios and spins. 

A 2 mg/cm$^{2}$-thick $^{24}$Mg target was bombarded
with $^{12}$C beams at E/A=53 and 95 MeV, with
an intensity of a few 10$^{7}$ pps, produced by the GANIL cyclotrons. The data presented in this work were collected with the Indra detector array \cite{pou95,pou96} and one of the forward rings of the Chimera 4$\pi$ detector \cite{chimera}. In peripheral collisions, $^{10}$C quasi-projectiles are largely produced. We reconstructed their internal unbound states by selecting those peripheral events where two alpha particles and two protons were detected in the forward part of the the detector setup. The longitudinal velocities, $v_{//}$, of these four particles were peaked around beam velocity, $v_{beam}$. Therefore, we used the condition $v_{//}>v_{beam}/2$ to select those events where 2$\alpha$+2$p$ particles can be unambiguously associated to the decay of primary $^{10}$C quasi-projectiles. More details about isolating $^{10}$C projectile decays can be found on Refs.     \cite{Gre06} and \cite{chb05}.

\begin{figure}
\centering
\includegraphics[scale=0.7]{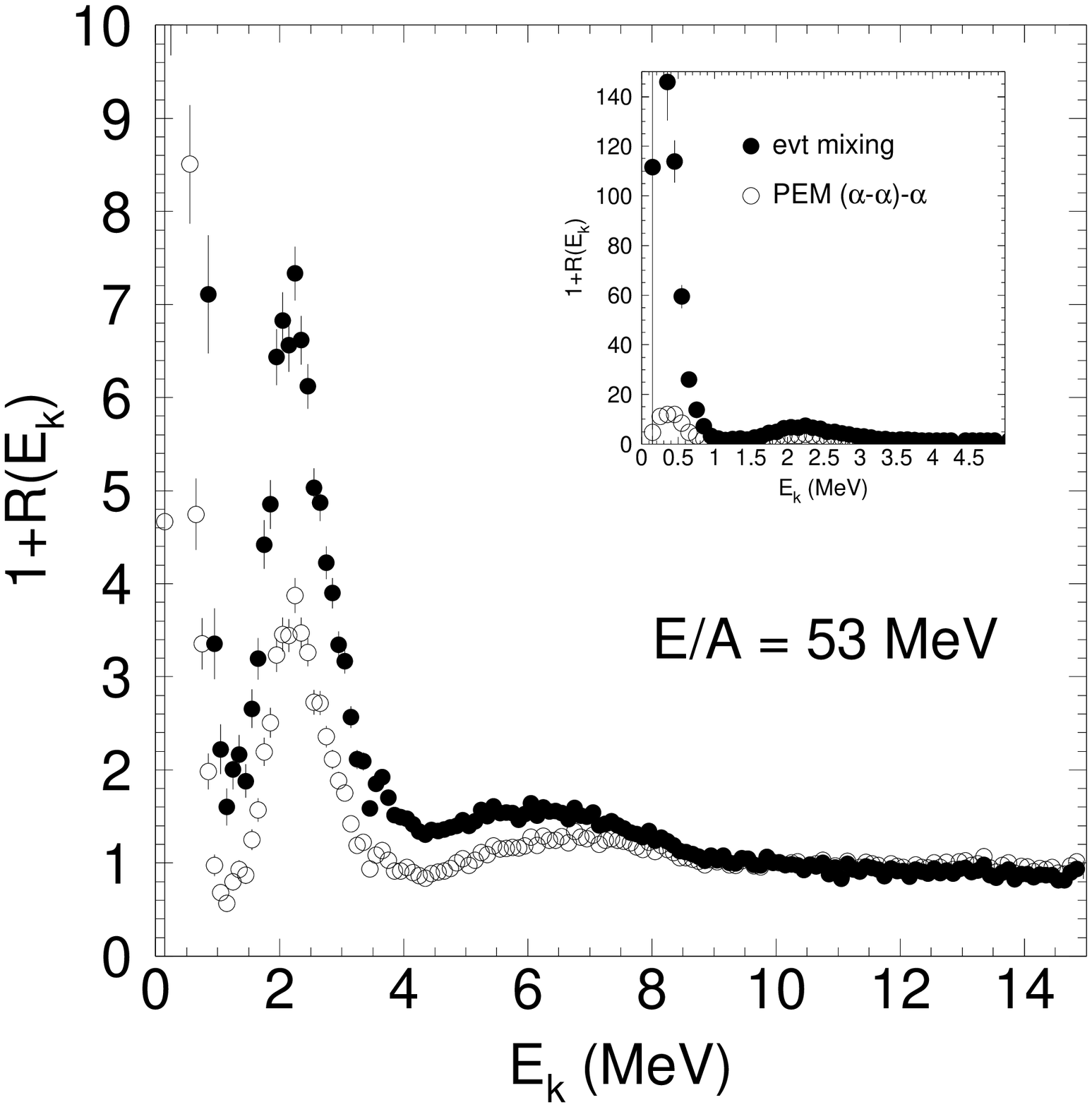}
\caption{Three-$\alpha$ correlation function. Full dots: 1+R(q) constructed with standard event mixing technique. Open symbols: 1+R(q) constructed with the partial event mixing technique (see text). An expanded view of the correlation function for $E_{k}\leq$5 MeV is shown in the inset.}
\label{corr12c}
\end{figure}

In order to illustrate our analysis techniques, we first study the most peripheral events where excited $^{12}$C quasi-projectiles are produced and are reconstructed by detecting three alpha particles with $v_{//}>v_{beam}/2$. Similarly to previous works \cite{cha95,ger06}, we constructed the three-alpha particle (3$\alpha$) correlation function:
\begin{equation}
1+R\left(E_{k}\right)=\frac{Y_{corr}(E_{k})}{Y_{uncorr}(E_{k})}
\label{eqfctcorr}
\end{equation}
In this equation, the correlated yield spectrum, $Y_{corr}(E_{k})$, is constructed with
3 $\alpha$ particles detected in the same event. This spectrum is sorted with respect to the quantity E$_{k}=\sum_{i=1}^{3}e_{ki}$, with $e_{ki}$ being the kinetic energy of the $i$-th alpha particle calculated in the center of mass of the three-body system. When a $^{12}C$ quasi-projectile is produced at an excitation energy $E^{*}$, it decays into the three alpha particles detected in coincidence. Then it follows that $E_{k}=E^{*}+Q$, where $Q$ is the mass-difference in the $^{12}C\rightarrow 3\alpha$ channel. 
The uncorrelated yield spectrum, $Y_{uncorr}(E_{k})$, is constructed with 3 $\alpha$ particles 
detected in three different events, by using the so-called \textit{event
mixing} technique~\cite{Zaj84,dri84}. The full dots on Fig. \ref{corr12c}
show the obtained 3$\alpha$ correlation function in $^{12}$C+$^{24}$Mg collisions at E/A=53 MeV. 
This correlation function allows one to explore the
internal states of $^{12}$C. The Q-value for the 3$\alpha$ decay
of $^{12}$C is -7.27 MeV. The first peak observed at $E_{k}\simeq$0.5~MeV
corresponds to the 7.65 MeV state in $^{12}$C. The second peak, centered
at $E_{k}\simeq$2.2 MeV, can be associated to the 9.65 MeV state,
while the third bump at $E_{k}\simeq$6 MeV results from
the overlap of highly lying closely packed states. The width of the observed
resonances is affected by the finite angular resolution of the apparatus.
Stimulated by the analysis technique described in Ref. \cite{cha95},
we modified the definition of the uncorrelated three particle yields
used in the denominator of Eq. \ref{eqfctcorr} by using a \textit{partial
event mixing} (PEM) technique. In particular, the open symbols in Fig. \ref{corr12c}
correspond to the correlation function obtained when the denominator
is constructed with two $\alpha$ particles taken from the same event
and the third one from a different event. The resonant peaks are still observed
in the correlation function but with a reduced magnitude. This
reduction can be partly attributed to the presence of 2-body correlations
associated to sequential decays of $^{12}$C. Indeed, the decay of
certain internal $^{12}$C states can proceed via the emission of $^{8}$Be+$\alpha$ pairs, with the very loosely bound $^{8}$Be
nucleus subsequently decaying into two $\alpha$ particles. These second chance 
$^{8}$Be decays are included in the numerator of Eq.\ref{eqfctcorr}
but not in the denominator, if the latter is evaluated with the standard
event mixing technique. In contrast, when the PEM technique is used, two $\alpha$ particles in the denominator are still taken from
the same event. Thus, some $\alpha-\alpha$ correlations due to secondary $^{8}$Be decays are kept in the denominator of Eq.~\ref{eqfctcorr}. Then, the $^{8}$Be decay contributions contained in the numerator of Eq.~\ref{eqfctcorr} are partially cancelled out by calculating the ratio, $Y_{corr}/Y_{uncorr}$, and the magnitude of the correlation function peaks is reduced. 
In this respect, the attenuation observed in the height of the correlation peaks demonstrates the existence of $^{12}$C sequential decay modes. 

\begin{figure}
\centering
\includegraphics[scale=0.55]{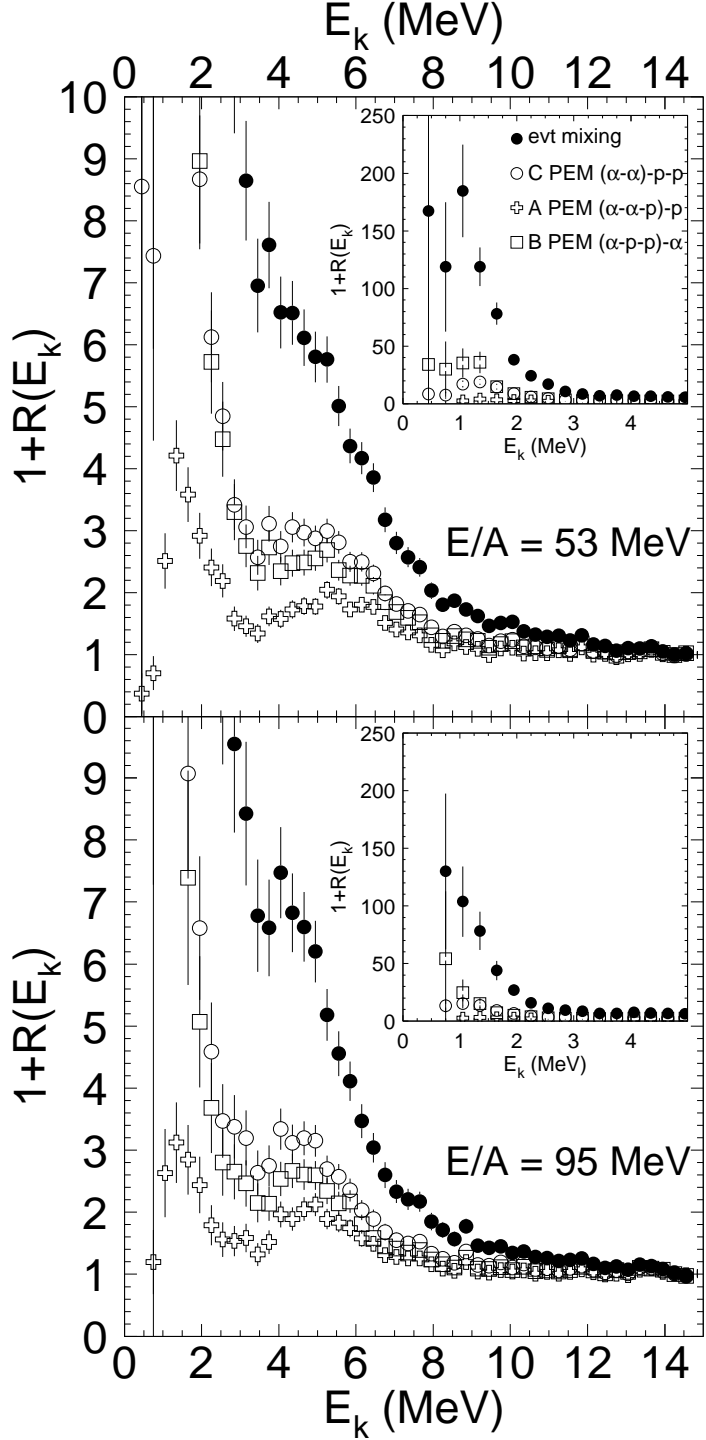}
\caption{2$\alpha$-2$p$ correlation functions at E/A=53 MeV (top panel) and 95 MeV (bottom panel). Solid symbols: standard event mixing technique. Open symbols: PEM techniques with two $\alpha$'s (circles), two $\alpha$'s and one proton (crosses) and one $\alpha$ and two protons (squares) taken from the same event (see text for details). Expanded views of the correlation functions for $E_{k}\leq$5 MeV are shown in the inset.}
\label{corr10c}
\end{figure}

We now turn to the study of $^{10}$C states by selecting those projectile breakup events where 2$\alpha$+2$p$ are detected in coincidence. 
Fig.\ref{corr10c} shows the four-particle 2$\alpha$-2$p$ correlation
function, $1+R(E_{k})$, constructed by using both the standard (closed symbols) and partial event mixing techniques described for the case of 3$\alpha$ correlation functions. 
The top (bottom) panel refers to E/A=53
MeV (E/A=95 MeV) incident energy. The 2$\alpha$-2$p$ correlation functions on Fig.~\ref{corr10c} show two broad peaks. The Q-value of the 2$p$+2$\alpha$ decay of $^{10}$C is -3.7 MeV. The first peak at $E_{k}\approx$1.5 MeV can
be associated to an overlap of $^{10}$C states at E$^{*}\approx$5.2-5.4 MeV. The second peak at $E_{k}\approx$5-5.5 MeV
corresponds to unknown states around E$^{*}\approx$9 MeV \cite{Wan93}. The location of
these peaks is independent of the beam energy, indicating that our access to intrinsic properties of the $^{10}$C system is not affected by possible artifacts induced by reaction dynamics. 

Regardless of the limited angular resolution of the INDRA detector, we attempt to extract information about the decay modes corresponding
to the intense peak at $E_{k}\approx$1.5 MeV. We search for signatures of sequential decays of $^{10}$C passing through the formation of its loosely bound subsystems $^{9}$B, $^{6}$Be and $^{8}$Be.
In particular, we explore the following decays sequences: A) $^{10}$C$\rightarrow^{9}$B$_{g.s.}$+$p$
with the $^{9}$B$_{g.s.}$ nucleus further undergoing
a decay $^{9}$B$\rightarrow\alpha$+$\alpha$+$p$; B) $^{10}$C$\rightarrow^{6}$Be$_{g.s.}$+$\alpha$ with the $^{6}$Be$_{g.s.}$ nucleus further decaying into
$\alpha$+$p$+$p$; C) $^{10}$C$\rightarrow^{8}$Be+$p$+$p$
with the $^{8}$Be$_{g.s.}$ nucleus further decaying into $\alpha$+$\alpha$. 

In order to search for contributions coming from the decay sequence A, we construct the denominator of Eq.~\ref{eqfctcorr} with a PEM technique where two $\alpha$ particles and one proton (2$\alpha$+$p$) are taken from the same event and the remaining fourth particle ($p$) is taken from a different event, obtaining the correlation functions shown as open crosses on Fig.~\ref{corr10c}. The magnitude of the first peak at E$_{k}\approx$1.5 MeV is strongly reduced to about 2$\%$ with respect to the case of complete event mixing (full dots). 
In line with what we have already discussed in the case of $^{12}$C, this result indicates that a strong contribution to the first
peak can be attributed to a decay of $^{10}$C proceeding
through the unbound $^{9}$B$_{g.s.}$ nucleus.
The decay sequence B, going through the formation of intermediate
$^{6}$Be$_{g.s.}$ nuclei is investigated using the PEM technique with one $\alpha$
particle and two protons from the same event ($\alpha$+2$p$) and the second $\alpha$
particle from a different event. In this case, the observed reduction of the peak magnitude (about 20$\%$ of the original peak height) is less pronounced than the one observed in case A. Finally, by selecting two $\alpha$
particles from the same event ($\alpha$-$\alpha$) and the two protons from different events, one explores the decay sequences of type C passing through
the formation of $^{8}$Be$_{g.s.}$ nuclei (open circles on Fig. \ref{corr10c}). The first peak is reduced to about 11$\%$ of the original peak height, indicating that the decay through $^8$Be$_{g.s.}$ is more likely than the decay sequence B but less important than decay sequence A. The stronger attenuation of the correlation peak height obtained with the PEM technique with 2$\alpha$+$p$ taken from the same event suggests a preference for the $^{10}$C states at E$^{*}$=5.2-5.4 MeV to decay through the intermediate formation of $^{9}$B$_{g.s.}$ nuclei, as compared to the decays through the formation of $^{8}$Be$_{g.s.}$ or $^{6}$Be$_{g.s.}$ nuclei. In particular, it seems that the decay sequence A is more likely than decay sequence B by a factor 10 and more likely than decay sequence C by a factor 5. 
 
In the case of the states of $^{10}$C corresponding to the peak observed at E$_{k}\approx$5-5.5 MeV, the attenuation of its magnitude observed when using the PEM techniques A, B and C 
is less pronounced as compared to the case of the lower lying state. This observation suggests that direct four body decays (2$p$+2$\alpha$) without passing through any intermediate unbound state are more likely for the states around E$^{*}\approx$9 MeV than the state for E$^{*}\approx$5.2-5.4 MeV. 
Among the studied sequential decay modes, sequence A is the most likely as it induces a peak magnitude attenuation of the order of 30$\%$ of its original value (see open crosses).  Decay sequences B and C seem
to provide comparable contributions to the decay of the studied $^{10}$C state, within statistical uncertainties, reducing the peak magnitude to about 35-40$\%$ of its original value. A slight systematic preference for B with respect to C is observed. In general, for both the first and the second peak we observe a preference of $^{10}$C to decay through the production of $^{9}$B$_{g.s.}$ nuclei rather than through the $^{8}$Be and $^{6}$Be systems. 

In order to confirm the $^{10}$C sequential decay modes, we study three-particle coincidence spectra for those events 
with 1 MeV$\leq$ E$_{k}\leq$ 3.5 MeV, corresponding to the first peak on Fig.~\ref{corr10c}. On Fig. 3 we show $\alpha-\alpha-p_{1}$ and $\alpha-\alpha-p_{2}$ kinetic energy spectra (top panels), $N(E_{k})$, where the kinetic energy, $E_{k}$, is calculated in the three-body reference frame. In particular, we choose the slowest ($p_{1}$) or the fastest ($p_{2}$) proton in the $^{10}$C reference frame. Similar spectra are constructed using the $\alpha_{1}-p-p$ and $\alpha_{2}-p-p$ coincidences (bottom panels on Fig. 3), where $\alpha_{1}$ and $\alpha_{2}$ particles are chosen using the same criteria as for $p_{1}$ and $p_{2}$. These spectra of Fig. 3 are expected to contain information about intermediate states in
$^{9}$B and $^{6}$Be nuclei eventually produced during the $^{10}$C decay sequences A and B described above. If $^{10}C^{*}$ did not decay through a definite sequence,
the $N\left(E_{k}\right)$ distributions would show no signatures of nuclear resonant decays. In contrast, we observe a peak at $E_{k}\approx$0.2 MeV in the $\alpha-\alpha-p_{1}$ spectrum, corresponding to the decay of $^{9}$B$_{g.s.}$ in the decay sequence A, and a peak at $E_{k}\approx$1.4 MeV in the $\alpha_{1}-p-p$ spectrum, corresponding to the decay of $^{6}$Be$_{g.s.}$ in decay sequence B. In the spectra $\alpha-\alpha-p_{2}$ and $\alpha_{2}-p-p$ (right panels on Fig. 3) we observe broad peaks possibly associated to the decay of higher $^{9}$B and $^{6}$Be excited states. 
These observations confirm our conclusions deduced from the PEM analysis shown on Fig.~\ref{corr10c}, providing a first  experimental evidence of sequential decay modes by unbound states of $^{10}$C exotic nuclei. More details about the analysis can be found in Ref. \cite{Gre06} and \cite{chb05}. Our estimates of contributions from different sequential decay modes have been recently confirmed by a dedicated experiment performed with a high resolution detector array and by using a different analysis technique aimed at exploring two-proton emission \cite{char07}.  

The results shown in the present work provide new insights about spectroscopic properties of $^{10}$C states. In particular, the existence of sequential decay modes and a semi-quantitative estimate of their contributions to the total width of $^{10}$C unbound states provide relevant information in order to determine their branching ratios and spins. Our results also show that heavy-ion collisions at intermediate energies can be used as a standard spectroscopic tool to access simultaneously important properties of several unstable nuclei that can be produced in one single reaction. This conclusion opens promising perspectives in collisions induced by radioactive beams where more exotic nuclear species are expected to be abundantly produced. 

In summary, we have used peripheral C+Mg collisions at intermediate energies to produce unbound $^{10}$C nuclei. By means of multi-particle correlation functions we access excited states in $^{10}$C and explore their decay modes. For the first time, an experimental evidence of sequential decays of $^{10}$C proceeding though the production of the intermediate loosely bound $^{6}$Be, $^{8}$Be and $^{9}$B nuclei is found. The used analysis techniques allow one to semi-quantitatively estimate the relative contributions of different sequential modes to the total width of the investigated states. A preference for a decay path producing unbound $^{9}$B nuclei is found. 

\begin{acknowledgments}
We thank the staff of the GANIL Accelerator facility
for their support during the experiment. This work was supported by
four France institutions : Le Commissariat \`{a} l'Energie Atomique,
Le Centre National de la Recherche Scientifique, Le Minist\`{e}re
de l'Education Nationale, and le Conseil R\'egional de Basse Normandie.
This work was also supported by the Natural Sciences and Engineering
Research Council of Canada, the Fonds pour la Formation de Chercheurs
et l'Aide \`{a} la Recherche du Qu\'{e}bec.
\end{acknowledgments}


\bibliography{reference}

\newpage

\newpage

\newpage

\begin{figure}
\centering
\includegraphics[scale=0.7]{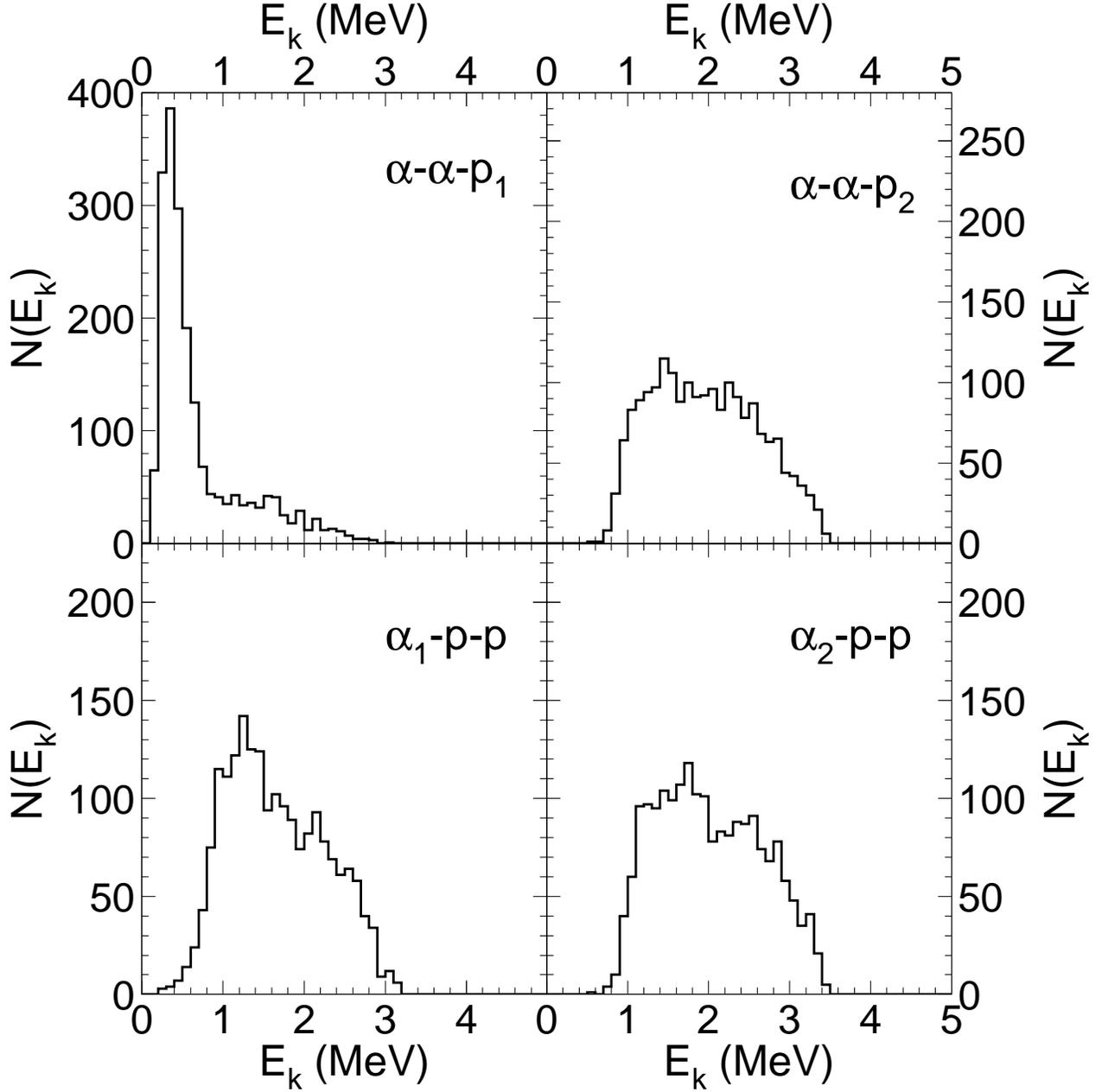}
\caption{Top panels: $\alpha-\alpha-p$ kinetic energy spectra constructed with the slowest ($p_{1}$, left panel) and the fastest ($p_{2}$, right panel) proton. Bottom panel: $\alpha-p-p$ kinetic energy spectra constructed with the slowest ($\alpha_{1}$, left panel) and the fastest ($\alpha_{2}$, right panel) $\alpha$ particle. The data correspond to an incident energy of E/A=53 MeV.} 
\label{seq10c}
\end{figure}

\end{document}